# The contribution of the ARIEL space mission to the study of planetary formation

D. Turrini[a,b,1], Y. Miguel[c], T. Zingales[d,e], A. Piccialli[f], R. Helled[g], A. Vazan[h], , F. Oliva[a], G. Sindoni[a], O. Panić[i,2], J. Leconte[j], M. Min[h], S. Pirani[k], F. Selsis[j], V. Coudé du Foresto[l], A. Mura[a], P. Wolkenberg[m,a]

[a] Istituto di Astrofisica e Planetologia Spaziali INAF-IAPS, Rome, Italy
[b] Departamento de Fisica, Universidad de Atacama, Copiapó, Chile
[c] Laboratoire Lagrange, Observatoire de la Côte dAzur, Nice, France
[d] University College London, London, UK
[e] INAF Osservatorio Astronomico di Palermo, Palermo, Italy
[f] Planetary Aeronomy, Belgian Institute for Space Aeronomy, Brussels, Belgium
[g] Institute for Computational Science, Center for Theoretical Astrophysics & Cosmology, University of Zurich, Zurich, Switzerland
[h] Anton Pannekoek Institute for Astronomy, University of Amsterdam, Amsterdam, Netherlands
[i] School of Physics and Astronomy, E. C. Stoner Building, University of Leeds, Leeds, UK
[j] Laboratoire d'Astrophysique de Bordeaux, Bordeaux, France
[k] Lund Observatory, Department of Astronomy and Theoretical Physics, Lund University, Lund, Sweden
[l] Observatoire Paris - Meudon, LESIA, Meudon Cedex, France
[m] CBK-PAN, Centrum Badań Kosmicznych Polskiej Akademii Nauk, Warsaw, PL

## Abstract

The study of extrasolar planets and of the Solar System provides complementary pieces of the mosaic represented by the process of planetary formation. Exoplanets are essential to fully grasp the huge diversity of outcomes that planetary formation and the subsequent evolution of the planetary systems can produce. The orbital and basic physical data we currently possess for the bulk of the exoplanetary population, however, do not provide enough information to break the intrinsic degeneracy of their histories, as different evolutionary tracks can result in the same final configurations. The lessons learned from the Solar System indicate us that the solution to this problem lies in the information contained in the composition of planets. The goal of the Atmospheric Remote-Sensing Infrared Exoplanet Large-survey (ARIEL), one of the three candidates as ESA M4 space mission, is to observe a large and diversified population of transiting planets around a range of host star types to collect information on their atmospheric composition. ARIEL will focus on warm and hot planets to take advantage of their well-mixed atmospheres, which should show minimal condensation and sequestration of high-Z materials and thus reveal their bulk composition across all main cosmochemical elements. In this work we will review the most outstanding open questions concerning the way planets form and the mechanisms that contribute to create habitable environments that the compositional information gathered by ARIEL will allow to tackle

**Keywords:** Atmospheric Remote-Sensing Infrared Exoplanet Large-survey; ARIEL; space missions; exoplanets; planetary formation; astrochemistry.

## 1. Introduction

The study of extrasolar planets brought our understanding of planetary systems and their formation to a turning point, showing us that our Solar System accounts for only a subset of all possible outcomes of the planetary formation process. Not only did the exoplanets highlight how migration is far more ubiquitous and plays a far more important role in shaping planetary systems than the current orbital structure of the Solar System suggests, but also revealed us the existence of types of planets that were previously just theoretical constructs. On the other hand, however, the exploration of the different planetary bodies of the Solar System has demonstrated that orbital and physical data (e.g., mass and radius) are intrinsically degenerate and, in many cases, can introduce a large degree of ambiguity in our understanding of the detailed nature and history of planetary bodies.

---

[1] Corresponding author (email: diego.turrini@iaps.inaf.it)
[2] Royal Society Dorothy Hodgkin Fellow

As a result, both pictures depicted by the Solar System and extrasolar planets are currently incomplete and potentially misleading, if considered individually. An illustrative example of this is offered by how these two fields of planetary science describe the family of planets. The classification used for the Solar System, when it comes to full-fledged planets according to the IAU definition, is based on their physical and compositional characteristics and distinguishes between terrestrial planets and giant planets, the latter further divided between ice giants and gas giants. This classification is incomplete as it makes a leap of an order of magnitude in mass encompassing the transition between terrestrial and giant planets, while we know that other planetary systems have super-Earths for example. Moreover, it implicitly makes assumptions on the composition/nature of planetary bodies in each category based on their Solar System analogues, like in the case of the ice giants.

The taxonomical classification generally used in the younger field of exoplanetary studies is, obligatorily, mostly phenomenological and, until recently, it mainly categorized planets based on their physical sizes (see e.g. Fressin et al. 2013). This classification includes Earth-sized planets (or simply "*Earths*"), *super-Earths*, *small Neptunes*, *large Neptunes*, and *giant planets* (encompassing both Jupiter-sized and super-Jovian exoplanets). This classification can also prove misleading: as we will discuss in more detail later, we know too little on the transition between large terrestrial planets, i.e. the super-Earths, and small giant planets, i.e. the small Neptunes. Planet size alone may not be a reliable indicator of the nature of these planetary bodies (see e.g. Spiegel et al. 2014), especially given the role planetary migration played in shaping the currently known population of exoplanets (see Sect. 2).

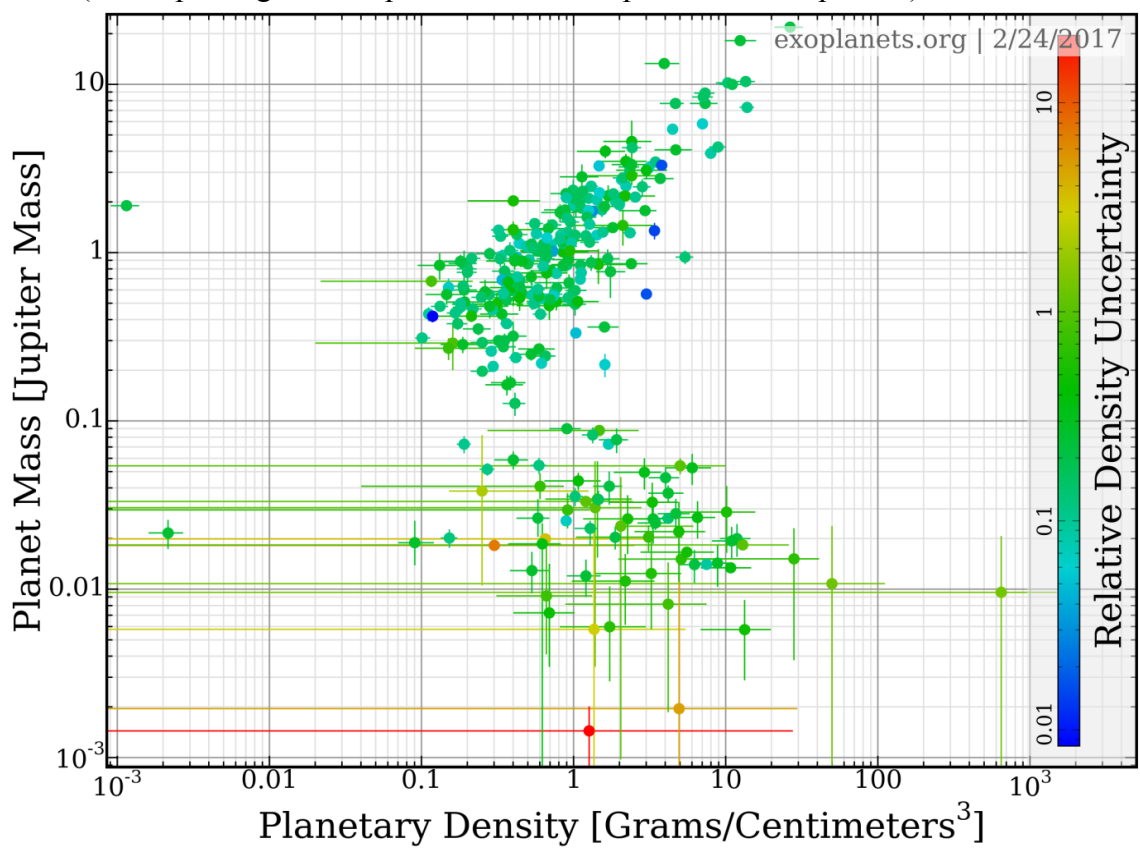

*Figure 1: Density-mass diagram showing the about 300 exoplanets for which we currently possess density estimates. The color bar reports the relative uncertainties on the density estimates on a logarithmic scale. Figure courtesy of www.exoplanets.org.*

A more reliable information on the physical and compositional nature of exoplanets is provided by their density, which to first order indicates the bulk composition and the gas/ice/rock ratios. A reliable estimate of the density requires the accurate knowledge of both mass and planetary radius: presently, we have density estimates for 10% of all currently known exoplanets and for most of these planets the relative uncertainty associated to such estimates is 20-50% (see Fig. 1, source: www.exoplanets.org). Relative uncertainties are particularly large when it comes to the densities of planetary bodies with masses below about 30 Earth masses (see Fig. 1), where only about one exoplanet out of two has a relative uncertainty below 30% (source: www.exoplanets.org). This makes it particularly difficult to discriminate, for planets characterized by similar radii, between water-poor and water-rich Earth-like planets or between gas-poorer super-Earths and gas-richer small Neptunes (see also Spiegel et al. 2014 for a discussion). Without this information it is difficult to derive reliable statistics or robust taxonomies.

The gathering of new and more accurate data on masses and radii in the coming years will improve this situation and provide a general picture of what exoplanets are made of. The experience with the planetary bodies of the Solar System, however, clearly tells us that this will be just the first step in a much longer journey. It is enough to compare the Earth with Venus or the dwarf planet Ceres with its kin Pluto, both pairs sharing very similar densities but being characterized by extremely different atmospheres and surface environments, to realize how limited is the information provided by density when it comes to fully characterizing a planetary body.

Specifically, due to the interaction with the surface and external environment (outgassing and/or accretion by asteroids and comets, changes in the obliquity, escape processes, and in the case of the Earth biochemical reactions), the atmospheres of the terrestrial planets in the Solar System underwent drastic changes evolving from a primordial to a secondary type. Earth and Venus were born as twins − formed at around the same time from the same ingredients and with a similar size. According to exoplanetary standards, Venus would be characterized as "Earth-like". Mixing of planetesimals and cometary impacts suggest that Venus and Earth may have also received a similar amount of volatiles. The amount of water on early Venus is estimated to have been equivalent to a global ocean between 5 and 500 m in depth (Baines et al. 2007). On present-day Venus the D/H ratio is 150 times larger than on Earth, indicating that the planet was wetter in the past but that the water was lost since then, mostly due to an intense runaway greenhouse (Donahue et al.,1982; de Bergh et al., 1991; Hunten, 1992).

Other mechanims imply that degassing, cometary impacts and escape processes, such as hydrogen escape may have played an important role in removing water from the atmosphere of Venus, removing more than a terrestrial ocean's worth of water during the first few hundred million years of Venus' evolution (Fedorova et al., 2008; Gillmann et al., 2009). The runaway greenhouse threshold is actually thought to be the key parameter in constraining the inner edge of the habitable zone around main sequence stars (Kopparapu et al., 2013) and is therefore fundamental to constrain the factors that produce it. In an alternative scenario for the atmospheric evolution of Venus, Gillmann et al. (2009) suggests Venus could have developed a dense molecular oxygen atmosphere (typically around 10 bar) formed by photolysis of water, with substantial amounts of water vapor. If Venus could keep a substantial amount of oxygen in its atmosphere for billions of years, an exoplanet similar to the young Venus would likely appear as a false positive of a planet hosting life (Gillmann et al., 2009).

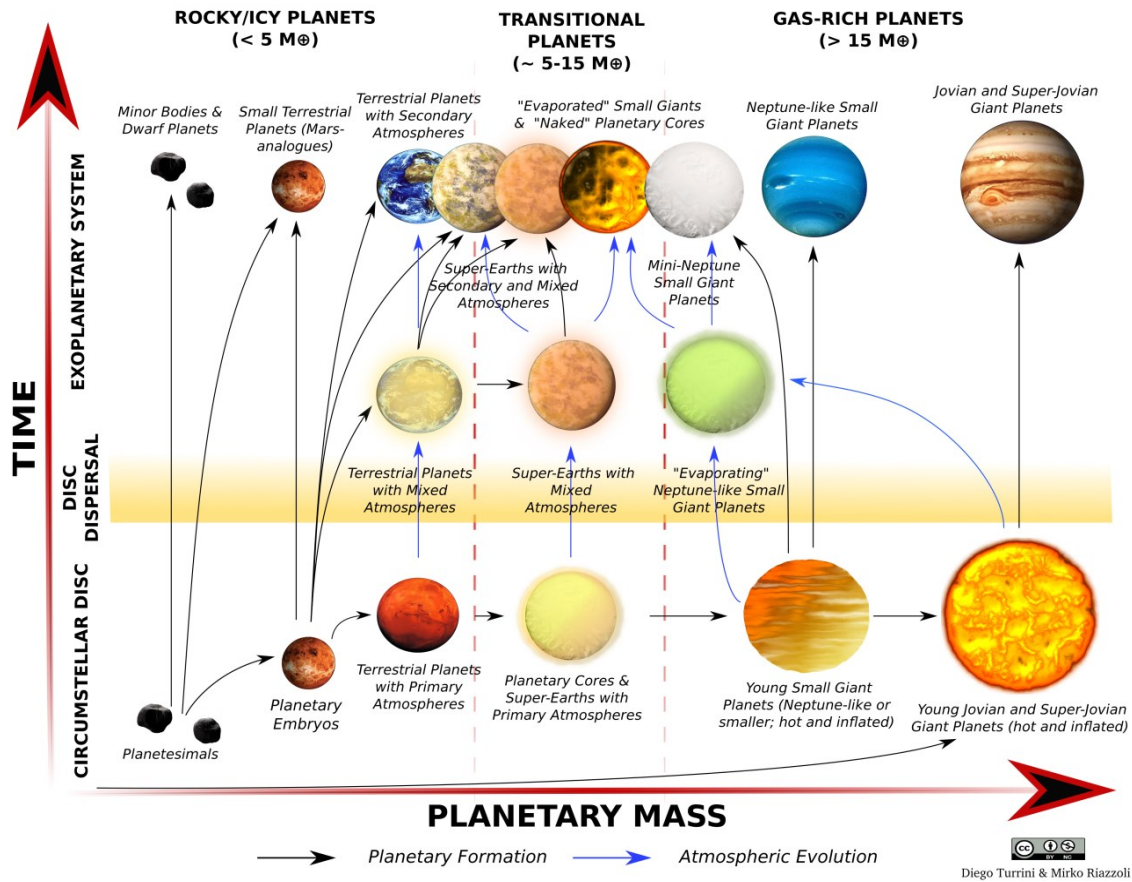

*Figure 2: Schematic representation of our current understanding of the formation and evolution paths responsible for the creation of the different kinds of planets currently known around the Sun and other stars. See main text for the definition of the three classes (dense, transitional and gas-rich) in which the different kinds of planets are grouped. Figure credits: Diego Turrini and Mirko Riazzoli.*

The lesson taught us by the Solar System is therefore that <u>to explore and understand the formation and evolution of a planetary body we need to characterize in detail its composition</u>. The lesson we learned from exoplanets is that <u>to grasp the extreme diversity of planetary bodies and planetary systems</u> existing in our galaxy <u>we need large and statistically representative samples</u>. The Atmospheric Remote-Sensing Infrared Exoplanet Large-survey (ARIEL) mission (Tinetti et al. 2017a,b), one of the three candidates as ESA M4 space mission, fulfills both these requirements as it will provide compositional information of hundreds of exoplanets. These data will thrust our knowledge forward and unveil the processes governing the formation and evolution of planets in our galaxy.

In the following sections we will discuss some of the ways in which ARIEL can change our understanding of planets and our place in the galactic context. Specifically, the scope of the following sections will be to highlight the most outstanding questions concerning the way planets form and the mechanisms that contribute to create habitable environments that ARIEL will allow to tackle. For more technical discussions of all the mechanisms that will be mentioned in the following, we refer the readers to the recent reviews by Morbidelli et al. (2012) and Raymond et al. (2014) regarding the formation of terrestrial planets, of D'Angelo et al. (2011) and Helled et al. (2014) regarding the formation of giant planets, to the insightful review by Morbidelli & Raymond (2016) for a discussion of open problems in the study of planetary formation and migration focusing both on exoplanets and the Solar System, and on the recent reviews by Massol et al. (2016) and Madhusudhan et al. (2016) for more details on the formation and evolution of planetary atmospheres and the link between their composition and planetary formation.

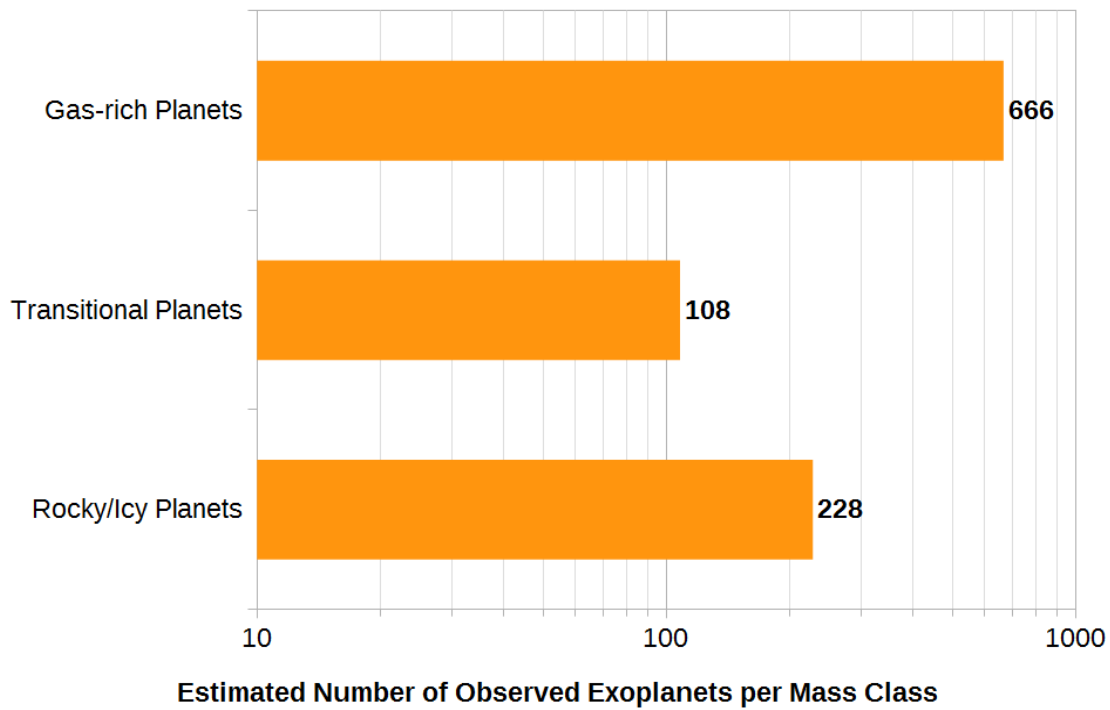

*Figure 3: One of the possible subdivisions of the ARIEL's observational sample among the three main classes of planets adopted in this document, highlighting how ARIEL will provide statistically significant samples for all classes. See main text and Zingales et al. (2017a,b) for further details.*

To avoid the pitfalls implicit in the taxonomies we discussed above, in the following discussion we will take advantage (for purely pragmatic reasons) of a simplified classification of planets. This classification, shown in Fig. 2, is based on our current understanding of how planets form and it groups planets on the basis of the time and environment of their formation. Time-wise (the vertical axis of Fig. 2), planets are divided according to whether they form before or after the dispersal of the circumstellar disc. The lifetime of circumstellar discs, which supplies the temporal reference frame of Fig. 2, varies between $10^6$ and $10^7$ years, with a median value of about $3 \times 10^6$ years (see Meyer 2008 and references therein and Fedele et al. 2010).

Mass-wise (the horizontal axis of Fig. 2), planets are simply divided into three main classes depending on their mass with respect to the critical mass range (5-15 Earth masses), for which theoretical models indicate that planetary bodies in a circumstellar disc can gravitationally accrete nebular gas to become giant planets (like their Solar System analogues Jupiter, Saturn, Uranus and Neptune).

Planets with masses below the critical mass range (i.e. those to the left in Fig. 2) will be mostly composed by mixtures of rock and metals or of rock, metals and ices (or, in terms of the most abundant cosmochemical elements, by Si, Mg, Fe, O and C, see Fegley & Schaefer 2010 and Spiegel et al. 2014). Due to the materials they are composed of, these *rocky/icy planets* are expected to be characterized by high densities. Planets with masses above the critical mass range (i.e. those to the right in Fig. 2) are expected to have accreted varying fractions of their mass as H and He from the circumstellar discs (see Spiegel et al. 2014). This broad category encompasses both planets dominated in mass by H and He (like Jupiter and Saturn) and planets possessing a massive envelope dominated by H and He but whose mass is mainly composed by rock and ices (like Uranus and Neptune). In the following they will be labeled as *gas-rich planets*.

Finally, planets whose mass falls within the critical mass range (i.e. those at the center on the horizontal axis of Fig. 2) will be labeled *transitional planets*. The planets falling into this category will belong to either the population of large super-Earths (i.e. rocky/icy planets with masses above about 5 Earth masses, according to the definitions given above) and to that of small Neptunes (i.e. gas-rich planets with masses below about 15 Earth masses according to the definitions given above), as the transition between these two families of planetary bodies should occur within this mass range (hence the label "transitional").

Until the discovery of exoplanets, the only regions of the diagram shown in Fig. 2 we could probe were those populated by the planets of the Solar System. This meant we could cover only the four corners of the diagram leaving the central region unexplored. Ground-based and space-based observational facilities are now allowing to probe with ever-increasing resolution the phases of the planetary formation process occurring during the life of circumstellar discs (sampling the bottom part of the diagram), while at the same time discovering more and more exoplanets (thus increasing the statistics of planetary bodies populating the upper part of the diagram).

ARIEL will complement our coverage of the phase space of this diagram by systematically exploring all the different types of planets available after the dispersal of the circumstellar discs down to terrestrial masses. As shown in Fig. 3, the observational sample of ARIEL will provide compositional information on hundreds of exoplanets in each of the three broad categories described above. Before proceeding to the discussion of how this information will advance our understanding, we need to introduce the process of *planetary migration* and illustrate the unique role it plays in making the science of ARIEL capable to transform our view of planetary formation. First, however, we will briefly summarize ARIEL's observational capabilities to provide context for the following discussion.

### 1.1. ARIEL's observational capabilities: an overview

ARIEL will use three different observing methods to investigated exoplanetary atmospheres: transit, eclipse and phase-curve spectroscopy (Tinetti et al. 2017a,b). The joint use of these methods allows to separate the signal of the planet at levels of 10-100 part per million (ppm) with respect to the signal of the star. Given the bright nature of ARIEL's targets, more sophisticated techniques such as eclipse mapping will support detailed investigations of the nature of their atmospheres (Tinetti et al., 2017a,b). Together with allowing to separate the signal of the planet from that of its host start, the use of these three observing methods will provide key information on the structure and variability of the exoplanetary atmospheres.

ARIEL will operate in a broad spectral range from 1.25 μm to 7.8 μm and will, additionally, take advantage of multiple photometric narrow bands in the optical range (Tinetti et al. 2017a,b). ARIEL's wide infrared spectral range includes many absorption and emission features of the gases mostly expected in exoplanetary atmospheres, e.g. $H_2O$, $CO_2$, $CH_4$, $NH_3$, HCN, $H_2S$, of the more exotic metallic compounds such as TiO, VO and of other species generally found in condensed state in the Solar System (Tinetti et al., 2017a,b). Together with exploring the atmospheric composition of exoplanets, the spectral capabilities of ARIEL will also allow to infer the properties of aerosols and clouds possibly present in these atmospheres.

The derivation of the relative abundances and the elemental compositions of the exoplanets from the spectra that ARIEL will measure can be achieved through spectral retrieval models. Retrieval tests showed that ARIEL will be able to retrieve several trace species from spectra having high signal-to-noise ratio (20) with an accuracy comparable to JWST (Tinetti et al., 2017a,b) if the molecular carriers have an atmospheric mixing ratio larger than $10^{-7}$ (Tinetti et al. 2017a,b). Moreover, Rocchetto et al. (2016) demonstrated that transit spectra recorded over a sufficiently broad infrared wavelength range can be effectively used to distinguish scenarios where the ratio between the two most abundant high-Z elements, C and O, is equal, larger or smaller than 1 (see also Sects. 2.1 and 3 for further discussion).

For further information on ARIEL observational capabilities and strategy we refer interested readers to Tinetti et al. (2017a,b) and to Barstow et al. (2017).

## 2. Planetary migration: ARIEL's best ally

The current sample of known extrasolar planets, even if biased toward more compact and/or more massive planetary systems than our own, highlights how planetary migration is a common and important process in shaping the structure of planetary systems. About half the exoplanets discovered so far orbit their host star at semimajor axes inferior to 0.1 au (source data: www.exoplanets.org). In the case of gas-rich planets (i.e. hot Jupiters and hot Neptunes) this is a strong theoretical indication that they should have formed elsewhere and migrated to their present position.

To provide an illustrative example, a circumstellar disc similar to the one assumed to have generated the Solar System (i.e. surface density at 1 au of 3000 g/cm$^2$ and scaling like $r^{-3/2}$, dust-to-gas ratio of 0.005 inside the water ice condensation line and 0.01 outside, and truncated at 0.1 au) would need almost all the mass contained between its inner edge and 2 au to produce a single planetary core with the minimal critical mass (i.e. 5 Earth masses) required to accrete gas. As the efficiency of the accretion process is likely much lower (a few 10%, see e.g. Chambers 2008) than this, the cores of most hot Jupiters and hot Neptunes should have formed farther away, likely beyond the water ice condensation line.

Migration can take place at different times in the life of a planetary system and can have different causes. It can occur during the life of circumstellar discs due to the exchange of angular momentum between the planet and the surrounding gas the evolutionary path indicated by the black and blue arrows in Fig. 4; (see e.g. Baruteau et al. 2016 and references therein), or after the disc dispersal as a results of the gravitational interactions between the different planetary bodies in the system (the evolutionary path indicated by the blue arrows in Fig. 4; see Weidenschilling & Marzari 1996, Ford & Rasio 1996). As a result, migration introduces multiple layers of degeneracy when investigating the nature and history of planets if only orbital data are considered. When we include the compositional data in the picture, however, the *widespread occurrence of migration and its capability of creating "hot" planets* (the end state of the evolutionary paths indicated by the blue and the black and blue arrows in Fig. 4), i.e. planets orbiting their host stars in regions characterized by equilibrium temperatures exceeding 1500 K, become *an important and powerful asset* for the same kind of investigations.

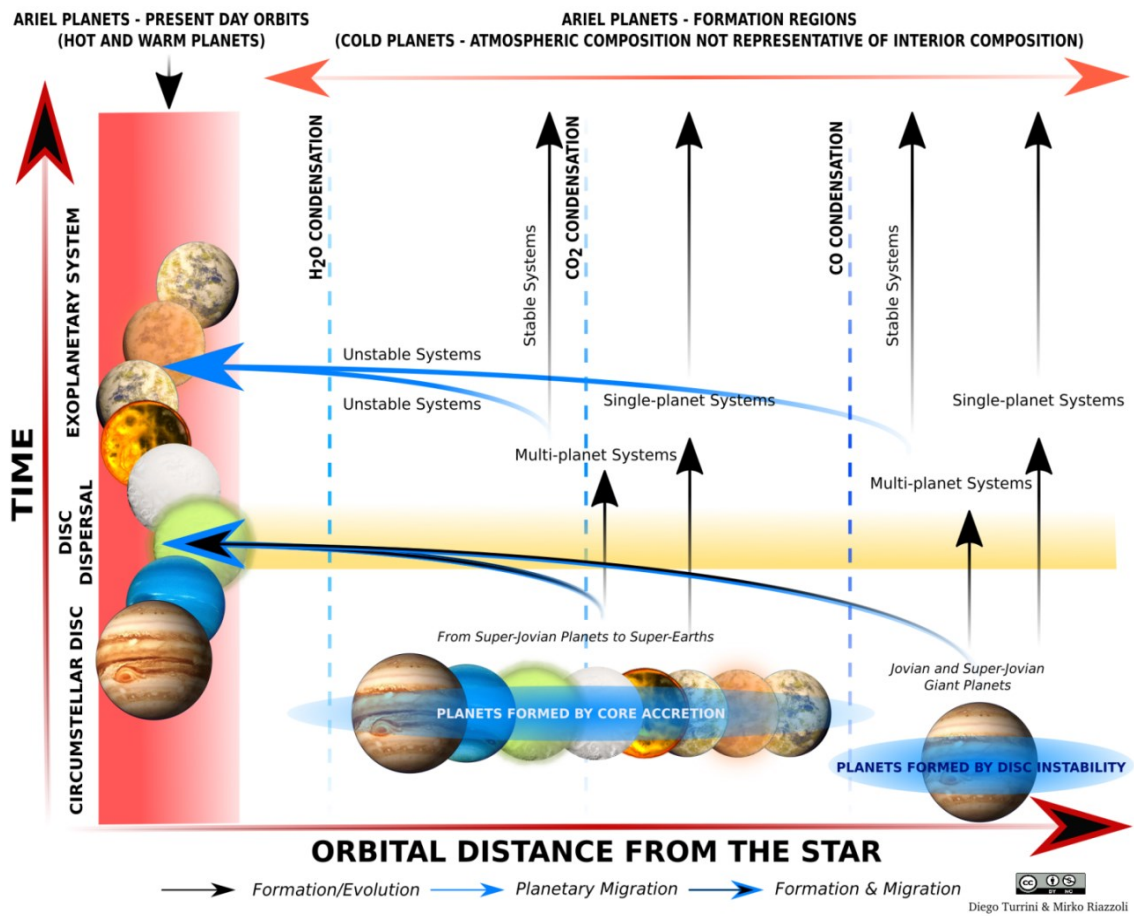

Figure 4: schematic illustration of the dynamical paths that can bring exoplanets formed through different processes in different, cold formation regions (where their atmospheric compositions is affected by condensation and removal processes) to the "hot" ($T_{eq} \geq 1500$ K) and "warm" ($T_{eq} \geq 1000$ K) orbital regions where ARIEL will observe them today (where their atmospheric composition is representative of their bulk composition) and of the different times at which they act. See main text for more details. Figure credits: Diego Turrini and Mirko Riazzoli.

The high temperatures experienced by *"hot" planets minimize the effects of condensation and atmospheric removal* of the less volatile elements (see e.g. Fegley & Schaefer 2010), making their atmospheres richer in information on their bulk composition than those of "colder" planets (i.e. those following the evolutionary paths indicated by the black arrows in Fig. 4 and remaining in the outer regions of their planetary systems). This is showcased by the compositional evolution of Jupiter's atmosphere across and after the impact of comet Shoemaker-Levy 9: the energy released by the impact temporarily allowed for the presence of more refractory elements and metals, which disappeared once the affected atmospheric regions cooled down (see Taylor et al. 2004 and Turrini et al. 2015 and references therein). The "hot" regions around stars therefore represent the optimal orbital locations of exoplanets for compositional studies.

Planetary migration delivers to these optimal locations planets that formed at different times, under different conditions and at different distances from their host stars, as schematically illustrated in Fig. 4 and discussed in more details for each of the three main classes of planets here considered in Sects. 3, 4 and 5. Once coupled with the large number of exoplanets ARIEL will observe (see Fig 3 and Tinetti et al. 2017a,b, Zingales et al. 2017a,b), this makes the *observational sample of ARIEL statistically complete from the point of view of the different formation and evolution tracks of the planetary bodies*. Since the way a planet forms and evolves and its interactions with the surrounding environment leave signatures in its bulk composition, the compositional information provided by ARIEL represents an unparalleled window into the process of planetary formation.

Planets as small as super-Earths can be affected by their interaction with the circumstellar disc and migrate on timescales shorter that the lifetime of the disc. Migration is relatively slow for planets with masses below or within the critical mass range, its timescale being roughly comprised between ~$0.5-1\times10^6$ years (D'Angelo et al. 2010, Baruteau et al. 2016). For planets with masses above the critical mass range, the migration timescales drops to a few $10^4$ years. Even in the slowest cases, however, such migration rates imply that planets with masses greater than a few Earth's masses can migrate from the outer regions of the disc to its inner edge well within the life of the disc itself.

Given the efficiency of disc-driven migration and the uncertainty (comparable to the lifetime of the discs themselves) on the timescale of formation of the different kinds of hot planets observed orbiting near their host star, planets orbiting at the same distances from their host star can come from a wide range of orbital regions (see Fig. 4). As a consequence, they would have formed under very different environmental conditions (i.e. beyond different ice condensation lines as illustrated inFig. 4) and therefore be characterized by very different elemental abundances (see Madhusudhan et al. 2016 and references therein for a discussion).

Disc-driven migration, however, is not the only mechanism that can be responsible for orbitally displacing a planet. Circumstellar disc, possessing orders of magnitude more mass than the planetary bodies, have a stabilizing effect on early planetary systems: after their disappearance, it is possible for the planets of multi-planet systems (currently about 40% of the known exoplanets are in systems containing two or more discovered planets, source: [www.exoplanets.org](www.exoplanets.org)) to find themselves on unstable orbital configurations (Weidenschilling & Marzari 1996; Ford & Rasio 1996). In such cases planetary migration will be the results of two-body and/or three-body effects (Weidenschilling & Marzari 1996; Ford & Rasio 1996) involving multiple planetary encounters with a chaotic exchange of angular momentum and energy between the bodies involved.

The process of chaotic exchange of angular momentum can result, among all possible outcomes, in the inward migration of one of the involved planets and the outward migration of another or in the inward/outward migration of one of the planets and the ejection of another from the planetary system. All these outcomes are generally associated with high final orbital eccentricities of the surviving planets (Weidenschilling & Marzari 1996; Ford & Rasio 1996). The first statistical studies of their orbital properties revealed the existence of an anti-correlation between the number of planets in planetary systems (i.e. a parameter also indicated as multiplicity) and their average eccentricities (Limbach & Turner 2015; Zinzi & Turrini 2017). Planets inhabiting systems with low multiplicity on average possess higher orbital eccentricities than planets in systems with high multiplicity, a trend holding from systems with multiplicity of two to the multiplicity of eight of the Solar System (Zinzi & Turrini 2017). This trend suggests that processes of chaotic exchange are widespread among multi-planet systems (Zinzi & Turrini 2017), supporting the idea that migration by planet-planet scattering plays an important role in shaping the structure of planetary systems.

## 2.1. Planetary migration, planetary composition and ARIEL

The elemental abundances in the bulk composition of a planet are first of all determined by its formation environment and its formation history, and only to a lower degree by its later evolution (see e.g. Turrini et al. 2015 and references therein). Planetary migration caused by the interactions with the circumstellar disc will act while the planet is still forming (the evolutionary path indicated by the black and blue arrows in Fig. 4), changing its surrounding environment and therefore affecting its final composition. Planetary migration due to multi-body dynamical effects will instead act after the planet already completed its formation process (see the evolutionary path indicated by the blue arrows in Fig. 4) and will not influence (or influence to a much lower extent) its bulk composition.

Depending on their dynamical evolution, therefore, "hot" planets that started their formation in the same environment or even at the same orbital distance from their host stars will show differences in their compositions. This multiplicity of formation histories and final compositions is already clearly highlighted by the sub-set of exoplanets for which we possess density estimates. As we show in Fig. 5 hot exoplanets, which can be identified with those orbiting a few hundredths of au from their host stars in the figure, possess a remarkable variability in terms of densities, which underlies an even greater variability in terms of compositions. This variability allows to tackle fundamental questions for all three classes of planets we defined in Sect. 1 and Figs. 2 and 3.

The interpretation of the compositional information and its link with the formation and evolution of planets is a rapidly evolving field of study. A growing body of work is exploring how the formation and migration of planets can be traced through the ratios of the most cosmically abundant atmophile elements C, O and N, with particular attention being devoted to the C/O ratio (see Madhusudhan et al. 2016 and references therein for a recent review). As an example, the C/O ratio of a "hot" planet can vary between 0.5 and 1, depending on whether the planet accretes gas both from beyond and interior to the water ice condensation line or if its gas accretion occurs only in the region beyond the water ice condensation line (see Turrini et al. 2015 and Madhusudhan et al. 2016 and references therein).

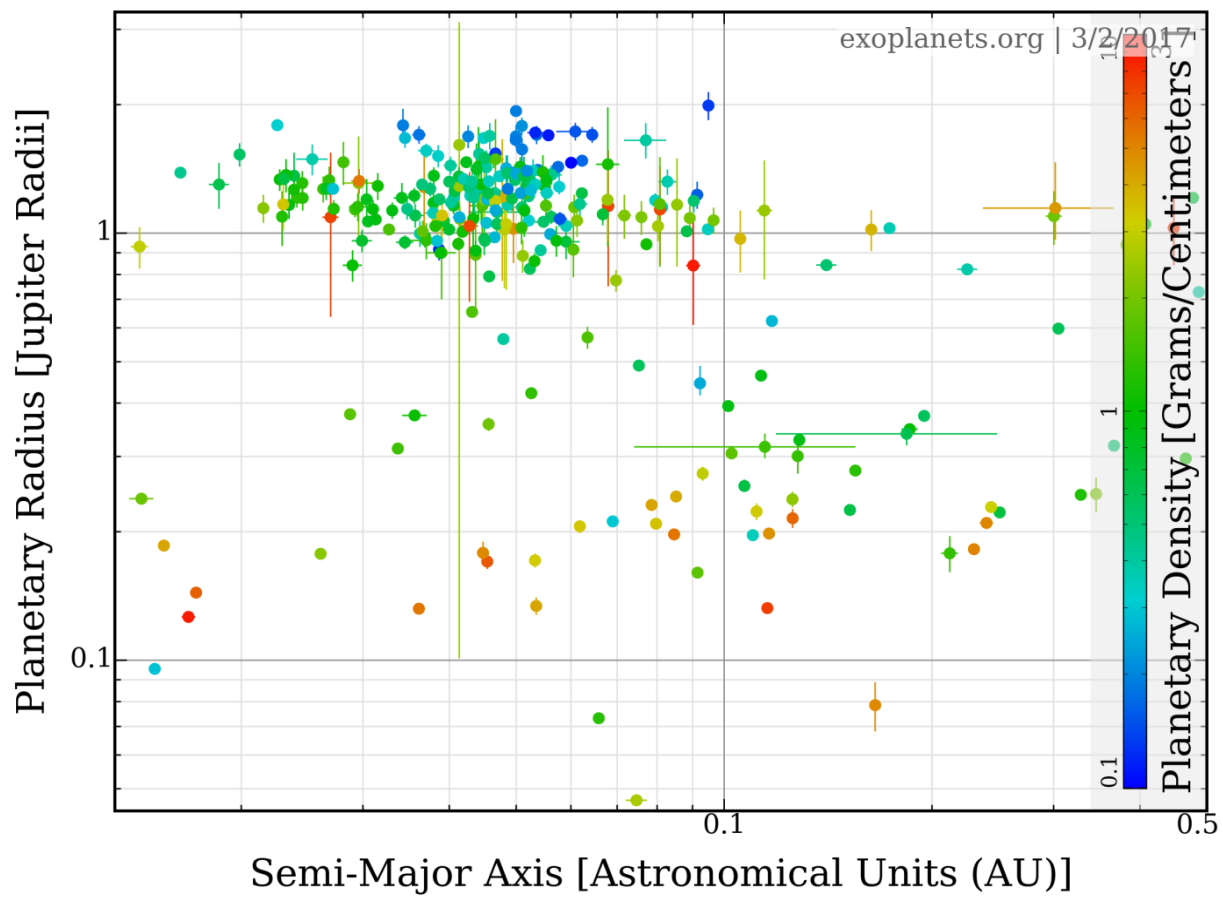

Figure 5: the compositional diversity of the currently known "hot" and "warm" exoplanets as revealed by their density (see colour bar). Figure courtesy of www.exoplanets.org.

It must be noted, however, that the C/O ratio alone (or, more generally, the elemental abundances of the atmophile elements mainly accreted with the nebular gas, see Fegley & Schaefer 2010) can provide misleading indications. As an example, in a recent paper, Espinoza et al. (2016) showed that the metal enrichment of the planetary envelope during the formation of giant planets can affect the C/O ratios. These authors estimated the atmospheric C/O ratio of nearly 50 relatively cool ($T_{eq}$< 1000 K) transiting gas giant planets and found C/O<1 in all cases, independently on the assessed formation locations, due to the contribution of the solid material accreted during the formation. Since low C/O ratios can lead to large water features, Espinoza et al. (2016) argued that water vapor absorption features should be ubiquitous in the atmospheres of metal-enriched transiting giant planets. Other factors that can affect the atmospheric C/O ratio of exoplanets are ultraviolet and X-ray irradiation, the atmospheric thermal structure and the strength of the transport processes (see Madhusudhan et al. 2016 and references therein).The spectral coverage of ARIEL and the large number of molecules that it allows to trace (see Sect. 1.1 and Tinetti et al. 2017a,b for details) will not only allow to verify the coupling between low C/O ratios and water features predicted by Espinoza et al. (2016), but will also allow to estimate the abundance of rock-forming, refractory and moderately volatile elements (see Fegley & Schaefer 2010 and references therein, Miguel et al. 2011). The information on these elements will allow to estimate the accretion of solid materials (both by rocky and icy planetary bodies) undergone by giant planets and to constrain in a more robust way their formation regions and dynamical evolution.

## 3. Exploring the nature of gas-rich planets with ARIEL

At the high end of the mass spectrum, the different classes of giant planets that we grouped together as gas-rich planets should form somewhere along the timespan covered by the lifetime of circumstellar discs (see Fig. 2), as the nebular gas of the discs is the only source of the H and He capable of supplying these planets with the bulk of their envelope mass and it is only available before the dispersal of the circumstellar discs. Two main processes have been proposed as possible pathway for the formation of giant planets (see Helled et al. 2014 and references therein): the core accretion scenario and the disc instability scenario (see Fig. 4).

In the *disc instability scenario* giant planets form as a result of a local gravitational instability in the circumstellar disc, which leads to the formation of a gravitationally bound object that collapses under its own self-gravity on timescales of the order of a few to a few tens of orbital periods. Planets formed by disc instability have enriched envelopes and can even acquire a core by sedimentation of the heavier elements present in the gas (and the dust trapped with the gas by the forming planet) and by accreting planetesimals just after their formation, though the actual efficiency of this process needs to be investigated further (see Helled et al. 2014 and Turrini et al. 2015 and references therein for a discussion).

The conditions requested for the nebular environment in the disc instability scenario favor the outer regions (few tens of au, see Fig. 4) of massive circumstellar discs during the earliest stages in the life of the latter as the formation environment of gas-rich planets. At such orbital distances, the short timescale of this process results in formation times of the order of or inferior to $10^4$-$10^5$ years. The disc instability scenario favors the formation of Jovian and super-Jovian giant planets, while the formation of planets similar to Uranus and Neptune, dominated by their planetary cores and with the gas representing less than 20% of their total masses, requires the forming planet to undergo significant mass loss from its envelope (see Helled et al. 2014 and references therein for a discussion).

The process that currently seems to best apply to the formation of gas-rich planets in the Solar System is the one described in the *core accretion scenario*, shown in Fig. 2 and also called *nucleated instability* scenario (see D'Angelo et al. 2010, Helled et al. 2014 and references therein). In the core accretion scenario, the gas-rich planets first form a planetary core of critical mass (i.e. ranging in mass between 5 and 15 Earth masses as mentioned before) by accumulation of solid material, meanwhile acquiring a more or less extended gaseous envelope by capturing gas from the circumstellar disc. When the mass of this expanded atmosphere becomes comparable with that of the planetary core, the gas becomes gravitationally unstable and begins to collapse on the core. This triggers a runaway gas accretion phase that causes a very rapid mass growth of the planet.

Depending on the actual size of the planetary core and on the amount of gas the forming planet can capture, the core accretion scenario can produce planetary bodies spanning the whole mass range covered by the gas-rich planets. The time required for the planetary core to reach the critical mass range and start accreting the gas should be of the order of a few $10^6$ years, while the runaway gas accretion timescale is quite shorter, ranging between a few $10^4$ years and a few $10^5$ years (see Helled et al. 2014). The formation time of gas-rich planets in the core accretion scenario is therefore dominated by the time required to form the critical-mass core. If this time is longer than the lifetime of the surrounding circumstellar disc, the core accretion process can also produce large super-Earths characterized by primary atmospheres captured from the nebular gas (see Massol et al. 2016 and references therein).

The two formation scenarios described above predict different formation times and formation environments for gas-rich planets, with different implications for their atmospheric and interior composition. In particular, the long time needed for the planetary cores to reach the critical mass in the nucleated instability scenario exposes the forming gas-rich planets to changes in the nebular environments in response to the evolving stellar radiation environment and to viscous heating. These changes will eventually result in variations in the relative composition of the gas and the solid materials that are accreted by the planet, as the ice condensation lines shift their positions throughout the discs (see e.g. Panić & Min 2017 for a recent discussion of this subject). Toward the end of their lives, it has also been suggested that dissipating circumstellar discs can experience a differential loss of the various gaseous species as a function of their molecular weight (Guillot & Hueso 2006), resulting in an additional change in the composition of the gas accreted by gas-rich planets.

Our capability to progress in our understanding of the formation of gas- rich planets is severely hindered by the fact *we have limited knowledge of their interior compositions*. This is partly due to our incomplete understanding of their formation process but is also due to the fact that the giant planets of the Solar System, our best template to study this class of planetary bodies, are *cold planets*. Their low temperatures make so that their atmospheric compositions, our only direct windows into their bulk composition, are extremely affected by condensation and removal processes. These processes remove the refractory and the less volatile species from the atmosphere, causing them to sink to depths beyond our probing capabilities.

As a result, while the enrichment in C of the four giant planets indicates a growing trend inversely proportional to the planetary mass (i.e. the less massive the planet, the more C-enriched with respect to solar abundance it is, see e.g. Hersant et al. 2004 and Atreya et al., 2016 and references therein) and, for the case of Jupiter, we know that also other volatile elements and noble gases are similarly enriched (see Atreya et al. 2016 and references therein), our knowledge of the overall metallicity of their gaseous envelopes is still affected by large uncertainties (see e.g. Guillot & Gladman 2000 and Miguel et al. 2016 for more recent results concerning Jupiter). In particular, we don't have a reliable picture of *how different elements contribute to the metallicity*: we don't know *whether all elements* within a given single gas-rich planet *are similarly enriched/depleted* with respect to the abundances of its host star, and we don't know *what is the rock/ice ratio* (i.e. the relative abundances of refractory and volatile materials) *of the high-Z component*. All these unknowns represent serious obstacles to identify the building materials of gas-rich planets when attempting to reconstruct their formation histories and constrain their formation environments.

By gathering compositional information on a large number of gas-rich planets formed at different distances from their host stars and delivered to the "hot" orbital regions (see Fig. 4), where the atmospheric composition is more representative of the internal one, ARIEL will allow for the *first reliable assessment of the interior compositions of gas-rich planets* for a statistically significant population (see Fig. 3). Moreover, the capability of ARIEL to observe elements belonging to all main cosmochemical groups (atmophile, moderately volatile, refractory and rock-forming elements, see Fegley & Schearer 2010 for a discussion) will allow to estimate the relative contributions of rocky planetesimals, icy planetesimals and gas to the metallicity of gas-rich planets.

An illustrative example of the compositional signatures that could be produced by the different formation and evolution histories of giant planets across the spectrum of elements that ARIEL will be able to observe is provided in Fig. 6. The cases shown in Fig. 6 were estimated under simplified assumptions and using a compositional model for the solid material based on the information provided by meteorites, asteroids and comets for the specific case of the Solar System (see Turrini et al. 2015 for further details) and, as such, should only be considered as qualitative examples. Within these limits, however, they showcase how the final C/O ratio of a gas-rich planet (Fig. 6, left panel) can change based on location of formation, dynamical history and also by the presence of other planetary bodies (e.g. the cases in the left panel of Fig. 6 where the planet accretes only gas, as the solids were assumed to have been depleted by a previously-formed planet, see Turrini et al. 2015 for details).

These cases also illustrate (Fig. 6, right panel) how the accretion of solid material from different orbital regions due to migration and the creation of three-body effects (orbital resonances, planet-planet scattering) can affect the relative abundances of elements outside the atmophile group (e.g. Fe/S, Cl/S, S/Ti). In principle, the accretion of solids from different orbital regions can also alter the ratios of elements belonging to the atmophile group (i.e C/O and N/O) due to the contribution of the O linked to Si, Mg and F in rocks. It should be noted, however, that the magnitude of these effects depends on the amount of solid material accreted by the gas-rich planet (see e.g. Turrini et al. 2015 and references therein): if the mass of solids is significantly smaller than that of gas, elemental ratios will be mainly determined by the composition of the gas (as illustrated also by the left panel in Fig. 6, where the changes of the C/O ratio caused by the accretion of solids are visibly smaller than those due to the formation region and the migration in the cases considering only gas accretion).

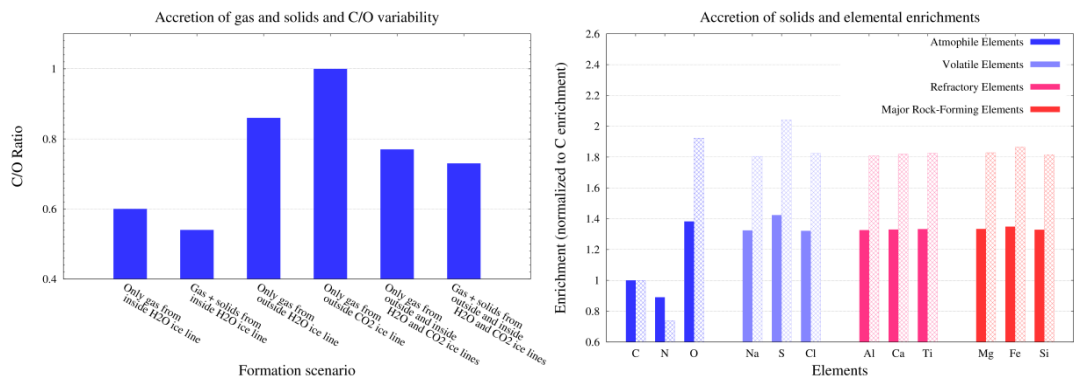

Figure 6: Left: examples of the effect of the formation and migration history of a giant planet on its atmospheric C/O from the simulations of Turrini, et al. (2015). Right: two examples of enrichment patterns created by the accretion of solids through the four major cosmochemical groups for elements that have spectral features in the observing bands of ARIEL, normalized to the enrichment in C produced by solids in each case. The solid bars on the left of each pair show the pattern created in a giant planet accreting solids mainly from beyond the water ice condensation line, the crisscrossed bars on the right of each pair show the pattern created in a giant planet accreting solids also from inside the water ice condensation line.

As the previous discussion highlights, numerous and often poorly constrained factors can play a role in affecting the final composition of gas-rich planets. The wide spectral range of ARIEL and the large sample of elements that the mission will be able to observe (see Sect. 1.1), however, offer a robust way tackle this problem. Rock-forming and refractory elements, in fact, give us information on the rocky component of the solid material accreted. Volatile elements are delivered by both rock and ices: with the information provided by the previous two classes of elements, we can disentangle the contribution of rock from that of ices. Atmophile elements are contributed both by solids and by gas: again, the information provided by the other classes of elements can help us disentangle the relative contributions of these two sources. This information, in turn, will allow for building a more robust picture of the environment(s) in which gas-rich planets form and, hence, on their formation process.

## 4. Exploring the nature of transitional planets with ARIEL

Moving toward smaller planetary masses, the second class of planetary bodies that will be investigated by ARIEL is the one that falls in the critical mass range. *This class of planets is the one we know the least* and the one that puts the more into question what we think we know about planetary formation as derived from the Solar System (see e.g. Morbidelli & Raymond 2016). Critical mass planets encompass both large super-Earths (large rocky/icy planets composed predominantly by the rock-forming elements Si, Mg, Fe, by O and possibly C, as defined in Sect. 1, and with atmospheres composed by high-Z elements, see e.g. Spiegel et al. 2014) and small Neptunes (small giant planets possessing an H/He-dominated envelope which however does not represent the bulk of their mass, see Spiegel et al. 2014) and one of the biggest open questions from a planetary formation point of view is where exactly the transition between these two populations occurs.

On one hand, according to our current theoretical framework, the formation of critical-mass planets should be able to occur during the lifetime of circumstellar discs to allow for these bodies to capture the nebular gas and become the planetary cores of gas-rich planets. On the other hand, there are no a priori reasons why the formation of super-Earths couldn't be a product of the same process governing the formation of rocky/icy planets (see Sect. 5) in which case, based on the chronological data from the case of the Solar System (see Morbidelli et al. 2012 and references therein), it would occur mostly after the dispersal of the gaseous component of the circumstellar disc.

Should the transition between one kind of critical mass planets to the other be relatively sharp, it would in principle be possible to reliably discriminate the two kinds of critical mass planets solely based on their mass-radius relationship. Within the limits imposed by their uncertainty, however, currently available data and theoretical studies of the process of atmospheric loss both suggest that the actual situation is more complex (see Massol et al. 2016 and references therein) and that the transition between Earth-like planets and Neptune-like planets is likely more continuous.

Planetary bodies reaching the critical mass range before the dispersal of the nebular gas will give rise to the exoplanetary population of small Neptunes, but under the right orbital and irradiation conditions (see Massol et al. 2016 and references therein) they may undergo significant atmospheric loss of hydrogen from their primary atmospheres, resulting in an metallicity increase. At the same time, super-Earths formed at the time of the circumstellar disc might capture a non-negligible mass of nebular H and He that, again under the right orbital and irradiation conditions, might survive the dispersal of the circumstellar disc. As such, the *planetary radius becomes a less and less reliable indication of the nature of the planet* in question the more we approach the region where the mass of the largest super-Earths overlaps that of the least massive small Neptunes (see also Spiegel et al. 2014).

Moreover, given that bodies in the critical mass range can already experience a significant migration due to their interaction with the disc (see e.g. D'Angelo et al. 2010, Baruteau et al. 2016 and references therein), also the information provided by the planetary mass and density can be misleading: a large, ice-rich super-Earth that formed farther away than the water ice condensation line and a small Neptune with a rocky-metallic core that formed nearer to the host star could in principle have quite similar densities (or, at least, similar enough to be difficult to discriminate within the measurement errors) despite their extremely different natures. The most reliable measure of the nature of a critical-mass planet would therefore be supplied by the composition of its atmosphere.

Planets in this transition region, therefore, are the ones for which the scientific observations of ARIEL can *transform the most our understanding of planetary formation*. Studying the transition between super-Earths and small Neptunes can be done to first order by estimating the abundances of the main atmospheric components and using the information provided by the mean molecular weight of the atmosphere to *estimate the abundances of hydrogen and helium* more reliably than the density alone would allow. These estimations of the atmospheric abundances of H and He allowed by ARIEL's observations for this class of planets will allow for more reliably estimating the relative populations of super-Earths and small Neptunes, and will directly and independently constrain the efficiency and timescale of the formation process of the planetary cores of gas-rich planets.

In addition to this important investigation, once properly discriminated, the samples of super-Earths and small Neptunes that ARIEL will observe in its exploration of the critical mass region will contribute to achieving the goals of the investigations targeted at gas-rich planets (see Sect. 3) and at rocky/icy planets (see Sect. 5), while at the same time providing an unprecedented set of data for the study of the atmospheric loss process.

## 5. Exploring the nature of rocky/icy planets with ARIEL

The formation process of rocky/icy planets is the one that in principle we know in more detail in the Solar System, in particular thanks to the radio-chronological data provided by the study of meteorites, planetary samples and terrestrial rocks. The formation of rocky/icy planets begins in circumstellar discs with the formation of the planetesimals, rocky and icy bodies spanning the size range characteristic of the present populations of asteroids and comets in the Solar System. The formation of the planetesimals has been proposed to occur throughout different processes (see Johansen et al. 2014 for a review) but we know that it takes place over a significant fraction of the life of the circumstellar disc (see Scott 2007 for an overview of the timescales of formation of different primordial bodies in the Solar System as derived from the radiometric study of meteoritic samples).

The largest planetesimals will evolve collisionally into planetary embryos, bodies conventionally defined as ranging in mass between 1% and 10% of the Earth's mass (i.e. roughly between the mass of the Moon and that of Mars), during the lifetime of the circumstellar disc, as testified in the case of the Solar System by the radiometric data provided by the meteorites originating from Mars (see Brasser et al. 2013 for a review). Planetary embryos, being formed during the lifetime of circumstellar discs, may initially possess tenuous primary atmospheres mainly composed of captured nebular gas. However, their mass is too low to prevent the loss of their initial primary atmospheres and their substitution with secondary atmospheres, as testified by the very case of Mars (see Massol et al. 2016 and references therein).

Theoretical models and radiometric data indicate that the formation process of Solar System's largest telluric planets, Venus and the Earth, took a few $10^7$ years to complete and therefore ended well after the dispersal of the circumstellar disc. Any primordial atmosphere possessed by the planetary embryos that concurred to their formation was likely lost by collisional ablation during the growth of these planets to their final masses and was substituted by secondary atmospheres generated by outgassing (see Massol et al. 2016 and references therein for a discussion of the involved physical processes)

Extrasolar planets, however, highlighted how the Solar System is not necessarily the most general or reliable reference case. On one hand, as confirmed by exoplanetary data the mass of the terrestrial planets in the Solar System does not represent any sort of physical limit: rocky/icy planets can easily range in mass up to a few times the Earth's mass, i.e. up to the lower boundary of the critical mass range. On the other hand, there is no a priori reason why, given the right conditions, rocky/icy planets should not reach their final mass during the lifetime of circumstellar discs and capture primary atmospheres from the nebular gas (see Massol et al. 2016 and references therein).

Depending on when a rocky/icy planet reaches a mass value that allows it to support a permanent atmosphere, therefore, its atmosphere could either be *primary* or *secondary* and the existence of Earth-sized and super-Earths with primary atmospheres should not be excluded a priori. In the case of initially primary atmosphere, moreover, outgassing processes linked to the geophysical evolution of the planet will increase the amount of heavier elements in the atmosphere, increasing its metallicity and creating a whole range of mixed atmospheres (analogously to what is created by atmospheric loss processes starting from small Neptunes, as discussed in Sect. 4).

The investigations of ARIEL of the atmospheric composition of rocky/icy planets will therefore allow to more reliably assess *the frequency of Earth-size planets and super-Earths with primary and mixed atmospheres* and expand our understanding of planetary accretion process beyond the constraints posed by the single case of the Solar System. In parallel, thanks to the exchange processes between the interiors, surfaces and atmospheres created by geophysical evolution (e.g. the outgassing from a magma ocean, see Massol et al. 2016 and references therein), the atmospheric observations of ARIEL will supply a wealth of detailed information on the interior compositions and the diversity of the population of hot, rocky/icy planets and, consequently, on their formation environment (e.g. inside or beyond the water ice condensation line).

### 5.1. Understanding the delivery of water to the habitable zone

The observational sample of rocky/icy planets to be observed by ARIEL will also allow to perform another investigation which holds the highest importance from the points of view of planetary formation and astrobiology: *understanding the mechanisms responsible for delivering water to the habitable zone*. The fact that the orbit of a dense planet is located in the habitable zone of its host planetary system does not guarantee that said planet will actually possess liquid water on its surface and in its atmosphere and will be, indeed, habitable. Since water condenses and is incorporated in solids at larger distances from the host star than those characteristic of the habitable zone (see e.g. Kopparapu et al. 2013), some mechanism is required to deliver it to the potentially habitable planets.

A substantial body of work on the origins of water on Earth (see e.g. Morbidelli et al. 2012 and references therein, Turrini & Svetsov 2014, Raymond & Izidoro 2017), the very definition of habitable environment, associate the presence of water on Earth to the presence of Jupiter in the Solar System. In general terms, this scenario identify in gas-rich planets a dynamical pathway for delivering water and volatile elements from beyond the water ice condensation line to the inner, dryer regions of a planetary system. The actual mechanism responsible for the delivery and the very time of the delivery are still debated (some works focus on the formation of the gas-rich planet, others on its migration and others yet on its secular dynamical sculpting of the planetary system) but the presence of a gas-rich planet is consistently a fundamental ingredient.

This scenario, if correct, would favor the presence of water in rocky/icy planets inhabiting multi-planet systems containing a gas-rich planet. This scenario, however, is not unique and a parallel body of work instead proposes that the presence of water on a rocky/icy planet can be unrelated to the presence of a gas-rich planet. Proposed mechanisms include the adsorption of nebular gas and water by dust grains (Drake 2005) and the accretion of water or hydrated materials through planetesimals migrating in the circumstellar disc (Quintana & Lissauer 2014), the latter mechanism actually being favored by the absence of gas-rich planets as it removes a dynamical barrier to the diffusion of hydrated materials toward the habitable zone. In this case, therefore, the presence and abundance of water on rocky/icy planets would likely be unrelated or anti-correlated from the presence of gas-rich planets.

As previously mentioned in Sect. 2, the ever-growing catalogue of exoplanets reveals that 40% of the currently known exoplanetary systems host two or more planets (source data: http://exoplanets.org). The recent discovery of the seven-planets system Trappist 1 (Gillon et al. 2017) and the information supplied by the dynamical properties of multi-planet systems (Limbach & Turner 2015, Zinzi & Turrini 2017) support the possibility that planetary systems characterized by high multiplicity like the Solar System are not uncommon in the Milky Way. By including in its observational sample terrestrial planets and super-Earths in both systems with and without the presence of one or more giant planets on outer orbits, ARIEL will allow to determine whether or not the presence of a giant planet results in a systematic difference in the atmospheric composition of the terrestrial planets. This finding, in turn, will provide an *unparalleled constraint to understand the mechanism(s) governing the presence of water in habitable planets and to understand which kind of planetary systems holds the best chances of hosting habitable environments.*

## Acknowledgements

D. Turrini and P. Wolkenberg gratefully acknowledge the support of the European Space Agency (ESA) during their participation to ARIEL's ESA Science Study Team. D. Turrini has been supported by the Italian Space Agency (ASI) under the contract 2015-038-R.0. The work of O. Panić has been supported through a Royal Society Dorothy Hodgkin Fellowship. A.Piccialli has received funding from the European Union's Horizon 2020 Programme (H2020-Compet-08-2014) under grant agreement UPWARDS-633127.